\documentclass[12pt]{iopart}
\usepackage{graphicx}
\usepackage{overpic}
\usepackage{amssymb}

\begin{document}
\title[]{Absolute instability modes due to rescattering of stimulated Raman scattering in a large nonuniform plasma}

\author{Yao Zhao$^{1,\dag}$, Zhengming Sheng$^{2,3,4,5}$, Suming Weng$^{2,3}$, Shengzhe Ji$^{1,6}$, Jianqiang Zhu$^{1,3}$}

\address{$^1$Joint Laboratory of High Power Laser and Physics, Shanghai Institute of Optics and Fine Mechanics, Chinese Academy of Sciences, Shanghai 201800, China}
\address{$^2$Key Laboratory for Laser Plasmas (MoE), School of Physics and Astronomy, Shanghai Jiao Tong University, Shanghai 200240, China}
\address{$^3$Collaborative Innovation Center of IFSA (CICIFSA), Shanghai Jiao Tong University, Shanghai 200240, China}
\address{$^4$SUPA, Department of Physics, University of Strathclyde, Glasgow G4 0NG, UK}
\address{$^5$Tsung-Dao Lee Institute, Shanghai Jiao Tong University, Shanghai 200240, China}
\address{$^6$University of Chinese Academy of Science, Beijing 100049, China}

\ead{$^{\dag}$yaozhao@siom.ac.cn}

\begin{abstract}
Absolute instability modes due to rescattering of SRS in a large nonuniform plasma are studied theoretically and numerically. The backscattered light of convective SRS can be considered as a pump light with a finite bandwidth. The different frequency components of the backscattered light can be coupled to develop absolute stimulated Raman scattering (SRS) and two plasmon decay (TPD) instability near their quarter-critical densities via rescattering process. The absolute SRS mode develops a Langmuir wave with a high phase velocity about $c/\sqrt{3}$ with $c$ the light speed in vacuum. Given that most electrons are at low velocities in the linear stage, the absolute SRS mode grows with much weak Landau damping. When the interaction evolves into the nonlinear regime, the Langmuir wave can heat abundant electrons up to a few hundred keV. Our theoretical model is validated by particle-in-cell simulations. The absolute instabilities may play a considerable role in the experiments of inertial confined fusion.
\end{abstract}


\maketitle

\section{Introduction}

Stimulated Raman scattering (SRS), the decay of incident laser into scattered light and electron plasma wave \cite{kruer1988physics,liu1974raman,Pesme2002Laser,Montgomery2016Two}, remains to be one of the major obstacles to direct-, indirect-, and possible hyper-drive schemes \cite{Betti2016Inertial,Lindl2014Review,Hurricane2014Fuel,Campbell2017Laser,He2016A}. SRS generates a large amount of hot electrons \cite{Smalyuk2008Role,Sangster2008High,Batani2014}, which can preheat the capsule. This has significant impacts on the implosion and ignition processes \cite{Dewald2010,Regan2010Suprathermal}. Even though there have been tremendous studies on the problem of hot electron production via SRS both in theory and experiments \cite{Montgomery2016Two,Glenzer2004Progress,Strozzi2008}, the generation mechanisms of hot electrons at nonlinear stages involving multiple driving waves are not yet fully understood.

Rescattering of SRS has been reported both theoretically and experimentally in different parametric conditions \cite{winjum2013anomalously,Kruer1980Raman,Hinkel2004National,Mima2001Stimulated,Klimo2010Particle}. Some previous works have discussed about the effects of SRS rescattering on the nonlinear scattering spectrum and saturation mechanism of instabilities in the regime where backward SRS is heavily damped \cite{Kruer1980Raman,Hinkel2004National}. Anomalously hot electrons due to rescatter of SRS have been studied recently \cite{winjum2013anomalously}. However, this work mainly studied the rescatter of SRS in homogeneous plasma, i.e., the absolute instabilities induced by rescattering in inhomogeneous plasma were not discussed. Generally, the absolute instability mode is the most significant instability in inhomogeneous plasma with the smallest threshold \cite{kruer1988physics,liu1974raman}. In addition, the phase velocity of the Langmuir wave excited by the absolute SRS is larger than other rescattering instability modes. In our work, we present a mechanism of hot electron production via the absolute instabilities in a large nonuniform plasma due to rescatter of SRS. It is well-known that parametric instabilities in inhomogeneous plasma are convective in one-dimensional (1D) geometry except near the quarter-critical density, where both SRS and two plasmon decay (TPD) become absolute instabilities \cite{liu1974raman,White1974Absolute}. For the indirect-drive scheme, even though there is a large underdense plasma region, the absolute instability is usually ignored since the maximum density of the plasma is typically less than 0.2$n_c$ \cite{myatt2014multiple}. In this work, we investigate the generation mechanism and associate conditions for the absolute instabilities. Our work suggests that the absolute instability modes induced via rescatter of SRS are the important mechanisms for hot electron production for long time interactions between laser and large scale inhomogeneous plasma. Our theoretical model is supported by particle-in-cell (PIC) simulations.

\section{Theoretical analysis of the absolute instability modes via rescatterering of SRS}

Here we consider cascaded scattering of SRS in an inhomogeneous plasma, including the first-order scattering occurring at the plasma electron density $n_1$, the subsequent second-order scattering at the density $n_2$, and the third-order scattering at the density $n_3$. The plasma has a positive density gradient along the laser propagation direction with an electron density range $[n_{min},n_{max}]$, where $n_{min}$ and $n_{max}$ are the minimum and maximum density, respectively. Assume that the corresponding electron plasma wave frequencies are given by $\omega_{p1}=\sqrt{4\pi n_1e^2/m_e}$, $\omega_{p2}=\sqrt{4\pi n_2e^2/m_e}$, and $\omega_{p3}=\sqrt{4\pi n_3e^2/m_e}$. According to the frequency match conditions for three wave coupling, the scattered light frequencies via the first-order scattering, second-order scattering and third-order scattering are given by $\omega_{s1}=\omega_0-\omega_{p1}$ (the corresponding wavenumber $k_{s1}\le0$), $\omega_{s2}=\omega_0-\omega_{p1}-\omega_{p2}$ (the corresponding wavenumber $k_{s2}\ge0$), and $\omega_{s3}=\omega_0-\omega_{p1}-\omega_{p2}-\omega_{p3}$ (the corresponding wavenumber $k_{s3}\le0$). The absolute instabilities in inhomogeneous plasma are found at the quarter-critical density of the incident electromagnetic wave. This can occur in certain regions in plasma via multiple SRS processes. Via multiple-order scattering, the incident laser energy is dissipated to plasma wave excitation and hot electron production.

\begin{figure}
\centering
    \begin{tabular}{lc}
        \begin{overpic}[width=0.6\textwidth]{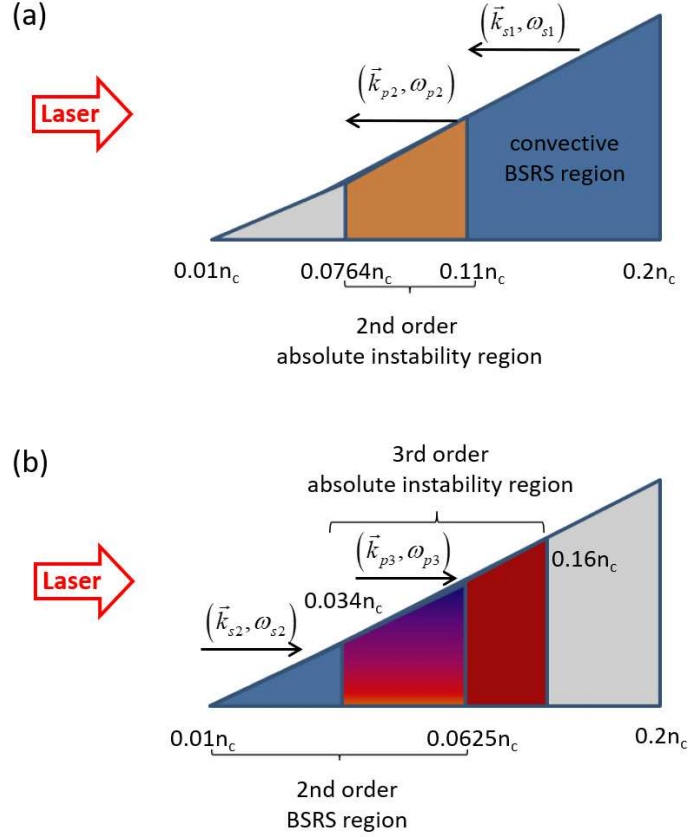}
        \end{overpic}
    \end{tabular}
\caption{ Schematic diagram for absolute instability regions due to (a) the second-order rescattering of SRS, and (b) the third-order rescattering of SRS in a linearly inhomogeneous plasma with density [0.01,0.2]$n_c$. BSRS means backscattering of SRS.
    }
\end{figure}

In the following, we examine the absolute instabilities via the second-order and third-order scattering. With the first-order backward SRS developed at $n_1$ and the second-order scattering at $n_2$, the second-order scattering becomes the absolute instability when it is developed at the quarter critical density of the first-order scattered light, i.e.,
\begin{equation}
\omega_{s1}=\omega_{s2}+\omega_{L2}\approx2\omega_{L2},
\end{equation}
where $\omega_{L2}=\sqrt{\omega_{p2}^2+3k_{L2}^2v_e^2}$ with $v_e$ the electron thermal velocity. In the following, we take $\omega_{L2}=\omega_{p2}$ for simplicity by ignoring the plasma electron temperature. Since $n_2\le n_1$, one finds that $n_1\ge n_c/9$ and $n_2\le n_c/9$ from Eq. (1). This indicates that the backscattering light from the region $n_1\ge n_c/9\approx0.11n_c$ can trigger absolute instabilities in the density area $n_2\le n_c/9$. One may note that the first-order forward SRS scattering light from $n_1\le n_c/9$ can induce absolute modes in the region $n_2\ge n_c/9$. Therefore, the major difference between rescatter of forward and backward scattering light is the density region for absolute instability. Without loss of generality, we mainly consider the rescatter of backward SRS in this paper.

With the condition $\omega_{s1}=2\omega_{p2}$ or $1-\sqrt{n_1/n_c}=2\sqrt{n_2/n_c}$ for the development of absolute instabilities, it is obvious that backscattered lights from different plasma density $n_1$ develop absolute modes at different density $n_2$. When the minimum quarter-critical density for the backscattering light $n_2^{min}$ is within $[n_{min},n_{max}]$, the absolute instability region is simply $[(1-\sqrt{n_{max}/n_c})^2/4, 1/9]n_c$. In this linear analysis, we ignored the coupling of backscattering lights from different plasma density between $n_c/9$ and $n_{max}$. As shown in Fig. 1(a), for example, the initial plasma electron density is [0.01,0.2]$n_c$, the minimum density for absolute SRS $n_2^{min}=0.0764n_c$ is found for backscattering from the density at $0.2n_c$. The first-order convective SRS occurs in [0.11,0.2]$n_c$, and its backscattering light can induce absolute instabilities by second-order rescattering of SRS within [0.0764,0.11]$n_c$.

Analogously, assume that the first-order backward SRS develops at $n_1$, and the second-order backward SRS develops at $n_2$, and the third-order backward SRS develops at $n_3$. If the third-order scattering is absolute, it is required that it is developed at its corresponding quarter-critical density, i.e.,
\begin{equation}
\omega_{s2}=\omega_{s3}+\omega_{p3}\approx2\omega_{p3}.
\end{equation}
Since $n_1\ge n_2$ and $n_3\ge n_2$, one finds that $n_2\le n_c/16=0.0625n_c$ according to Eq. (2). Therefore, the second-order backward SRS developed in $[n_{min},0.0625]n_c$ can induce the third-order absolute instability modes in a slightly higher density region. Note that $\omega_0=\omega_{p1}+\omega_{p2}+2\omega_{p3}\le\omega_{p1}+3\omega_{p3}$ and that $\omega_0=\omega_{p1}+\omega_{p2}+2\omega_{p3}\ge2\omega_{p2}+2\omega_{p3}$, one finds the absolute instability range of the third-order rescattering $[(1-\sqrt{n_{max}/n_c})^2/9,(1-2\sqrt{n_{min}/n_c})^2/4]n_c$. The absolute instability region may be reduced by the upper or lower limits of the plasma density profile. As an example, for the given plasma electron density shown in Fig. 1(b), the secondary backward SRS is developed in [0.01,0.0625]$n_c$, which can induce the third-order absolute instabilities in [0.034,0.16]$n_c$. Note that even though the absolute instability region via the third-order scattering is relatively larger than that via the second-order scattering, its intensity is much weaker. One finds a common region [0.0764, 0.11]$n_c$ for both the second-order and third-order absolute instability modes.

It is worthwhile to check with the phase velocity of the excited electron plasma wave, which determines the energy of trapped electrons. Without loss of generality, the wavenumber match condition for the second-order absolute SRS is $k_{p2}=k_{s1}-k_{s2}$. Since $k_{s2}\approx0$ for the absolute SRS, one obtains ${k}_{p2}\approx{k}_{s1}=-\sqrt{\omega_{s1}^2-\omega_{p2}^2}/c$. With the help of Eq. (1), one finds the phase velocity of the electron plasma wave is simply
\begin{equation}
{v}_{ph2}=\omega_{p2}/{k}_{p2}\approx-c/\sqrt{3}=-0.58c.
\end{equation}
The negative sign means that the phase velocity is opposite to the incident laser propagation as shown in Fig. 1(a). Corresponding to this phase velocity, the electron energy is about 170 keV. Similarly one can find the phase velocity of the electron plasma wave excited via third-order absolute SRS ${v}_{ph3}=\omega_{p3}/{k}_{p3}\approx c/\sqrt{3}$.

Now we discuss the necessary conditions for the development of absolute modes besides the density scale. Generally backward SRS is much stronger than forward SRS due to its larger growth rate. However, when the backward SRS is heavily Landau damped in a very hot plasma, the absolute instabilities will be mainly induced via the rescatter of forward scattering \cite{Hinkel2004National,Langdon2002Nonlinear}, which may play a role in the direct-drive scheme. Here we consider a plasma with a few keV where the backscattering SRS has not been heavily damped. The Landau damping of absolute SRS is weak due to the high phase velocity of the Langmuir wave. For example, assume the electron temperature is 2keV, i.e., thermal velocity $v_{th}=0.0626c$, which is a ninth of the Langmuir wave phase velocity $v_{ph}=0.58c$. The high temperature mainly reduces the saturation level of backscattering light, which acts as a damping of pump wave to develop absolute SRS. Therefore, the threshold for the development of absolute SRS induced via rescattering is mainly determined by the density scale length $L$ and incident laser intensity. As we known, the threshold for the first-order absolute SRS in inhomogeneous plasma is $a_0\gtrsim(k_0L)^{-2/3}$ \cite{liu1974raman,White1974Absolute}.
We treat the backscattering light as a pump wave, then we have $a_{s0}\gtrsim(k_{s0}L_s)^{-2/3}$ where $L_s=L\omega_s^2/\omega_0^2$ and $k_{s0}$ is the wavenumber of the first-order incident light, i.e., $k_{s0}=k_0$. By normalizing $a_{s0}$ back to the frequency of incident laser, we obtain the threshold for absolute SRS induced via second-order SRS rescattering
\begin{equation}
a_s\gtrsim\left(\frac{\omega_0}{\omega_s}\right)^{1/3}\left(\frac{1}{k_0L}\right)^{2/3}.
\end{equation}
Assume that $\omega_s=0.668\omega_0$ at $n_e=0.11n_c$, and $L=1000\lambda$, we obtain $a_s\gtrsim0.0037$. The threshold for incident light amplitude is $a_0\gtrsim0.013$ provided the SRS reflectivity is 8\% \cite{Fern2000Observed}. Note that the threshold for convective backward SRS at $n_e=0.11n_c$ is $a_0\gtrsim0.0136$ \cite{liu1974raman}, which indicate that as long as the reflectivity of convective SRS in $n_e>0.11n_c$ is larger than 8\%, the absolute SRS will eventually be induced.

\begin{figure}
\centering
    \begin{tabular}{lc}
        \begin{overpic}[width=0.6\textwidth]{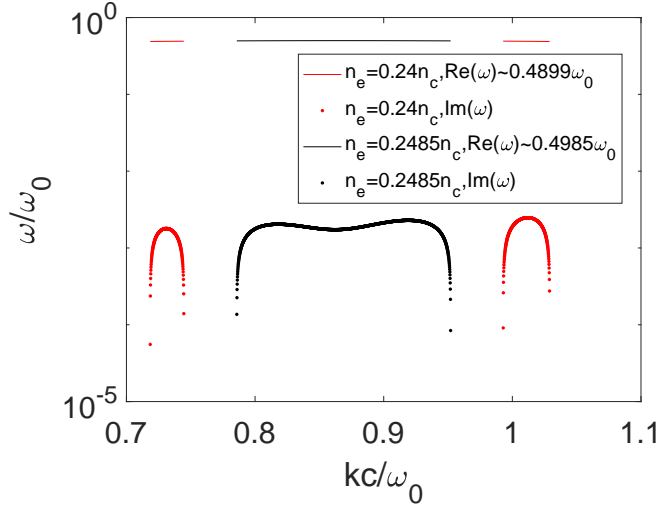}
        \end{overpic}
    \end{tabular}
\caption{ Numerical solutions of SRS dispersion equation at the plasma density $n_e=0.24n_c$ and $n_e=0.2485n_c$. The dotted line and continuous line are the imaginary part and the real part of the solutions, respectively.
    }
\end{figure}

The above linear analysis only considers individual scattering at local plasma density. When considering the absolute SRS instability, waves scattered at different plasma densities with different frequencies can be strongly coupled. Let us consider the bandwidth effects of backscattering light at the near-quarter-critical density, where the mismatch of wave numbers becomes much weak. The width of the instability region in the plasma wave vector $\Delta k$ is proved to be a critical factor for the modes coupling \cite{YaoZ2017Effective}. The dispersion for SRS in homogeneous plasma is given by \cite{kruer1988physics}
\begin{equation}
\frac{\omega_p^2a^2_0}{4}\left(\frac{c^2k^2}{D_e}+1\right)\left(\frac{1}{D_+}+\frac{1}{D_-}\right)=1,
\end{equation}
where $D_e=\omega^2-\omega_p^2, D_\pm=\omega^2-k^2c^2\pm2(\omega_0\omega-k_0kc^2)$.
Numerical solutions of Eq. (5) at different densities with the normalized laser amplitude $a_0=0.01$ are plotted in Fig. 2. The relation between $a_0$ and laser intensity $I_0$ is given by $a_0=\sqrt{I_0 (\mathrm{W}/\mathrm{cm}^2)[\lambda(\mu \mathrm{m})]^2/1.37\times 10^{18}}$. As show in Fig. 2, at $n_e=0.24n_c$, the instability regions of forward and backward SRS are well-separated in their wave vectors with almost equal width. However, when the plasma density increases to very near the quarter-critical density $n_e=0.2485n_c$, one finds the forward and backward SRS sharing a common plasma wave. The width of the coupling instability region can be estimated as twice of the backward instability region at $n_e=0.2485n_c$, i.e., $\Delta kc\approx16.8a_0\omega_0=0.168\omega_0$. Such a large instability region can greatly increase the coupling of light beams with different frequencies for the development of absolute SRS instabilities \cite{YaoZ2017Effective}. Therefore, in inhomogeneous plasma, even if the first order backscattering light may have broad bandwidth, they can be strongly coupled in their subsequent scattering for absolute SRS. Note that when the absolute SRS is developed, the scattered light waves have the wave number $k_s\approx0$. The scattering light at the quarter-critical density can be described by near-zero refractive index photonics, which contains rich physics and extensive applications \cite{Liberal2017,Ahmedm2014}. The trapped light with $k_s=0$ may be one of the reasons for laser energy deficit in inertial confined fusion \cite{Klimo2010Particle}. One notes that the broad width of $k$ spectrum also indicates a wide range of the Langmuir wave phase velocity, and the median of which is given by Eq. (3). A large range of phase velocity leads to a staged acceleration of electrons.

The coupling mechanism of TPD modes is similar to the SRS case. As known from the growth rate of TPD \cite{kruer1988physics}
\begin{equation}
\Gamma\simeq\frac{c\mathbf{k}\cdot\bf{a}_0}{4}\left|\frac{(\mathbf{k}-\mathbf{k}_0)^2-k^2}{k|\bf{k}-\bf{k}_0|}\right|,
\end{equation}
the width of TPD instability region is much larger than the one of SRS. Therefore, the scattering lights can be coupled to develop TPD in the absolute instability region.

In conclusion, the linear analysis suggests that a laser propagating in large inhomogeneous plasma can generate a large region of absolute instabilities by cascaded scattering and coupling of different frequency components of scattering light. Both the second- and third-order rescattering of SRS contribute to the development of absolute instabilities. Even though the intensity of the third-order scattering is relatively weak, it can still be strong enough to heat abundant electrons in a long time interaction.

\section{Simulations for absolute instabilities due to the second-order rescattering of SRS}
\subsection{1D simulations for absolute SRS mode}

To validate the above theoretical prediction, a series of PIC simulations have been performed in different plasma density region by using the {\sc klap} code \cite{chen2008development}. Firstly, one-dimensional (1D) PIC simulations have been carried, which avoid mixing-up of various instabilities and enable one to identify the development of the multiple SRS rescattering instabilities more clearly. The space and time given in the following are normalized by the laser wavelength in vacuum $\lambda$ and the laser period $\tau$. The length of the simulation box is 700$\lambda$, where the plasma occupies a region from 25$\lambda$ to 650$\lambda$ with density profile $n_e(x)=0.08[1+(x-25)/1000]n_c$, with $x$ the longitudinal axis. To exclude the third-order absolute mode, the minimum plasma density is set at 0.08$n_c$, and the maximum density is 0.13$n_c$. The initial electron temperature is $T_{e0}=100$eV. The ions are stationary with a charge $Z=1$. Reflection boundary conditions have been set for the electrons. A linearly-polarized semi-infinite pump laser with a uniform amplitude $a_0=0.02$ (the corresponding intensity is $I_0=4.5\times10^{15}\mathrm{W}/\mathrm{cm}^2$ with $\lambda=0.35\mu m$) is incident from the left boundary of the simulation box. We have taken 100 cells per wavelength and 50 particles per cell.

\begin{figure}
\centering
    \begin{tabular}{lc}
        \begin{overpic}[width=0.83\textwidth]{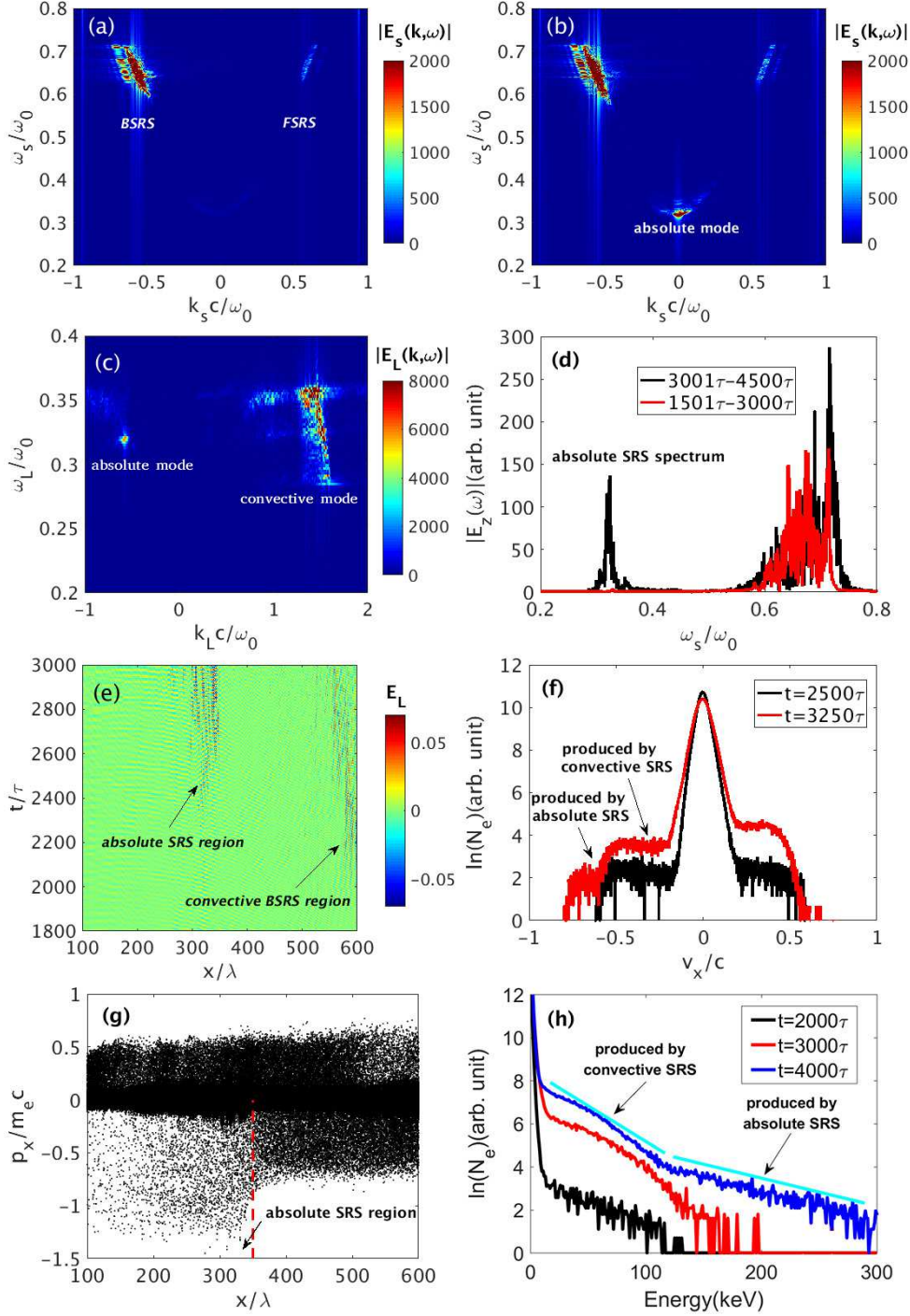}
        \end{overpic}
    \end{tabular}
\caption{ PIC simulation results for the development of the absolute SRS via the second-order scattering. (a) and (b) Wavenumber-frequency distributions of the scattered light in the time windows $[1501,2000]\tau$ and $[2001,2500]\tau$, respectively. FSRS means forward scattering of SRS. (c) 2D Fourier transform $|E_L(k,\omega)|$ of the electric field in the time windows $[2001,2500]\tau$. (d) Fourier spectra of backscattered light diagnosed at $x=10\lambda$. (e) Time-space distributions of Langmuir waves, where $E_L$ is the longitudinal electric field normalized by $m_e\omega_0c/e$, $m_e$, $c$ and $e$ are electron mass, light speed in vacuum, and electron charge, respectively. (f) Longitudinal velocity distributions of electron at different time. (g) Longitudinal phase space distribution of electrons near the region of the absolute SRS instability at $t=3250\tau$. (h) Energy distributions of electrons at different time, where $N_e$ is the relative electron number.
    }
\end{figure}

\begin{figure}
\centering
    \begin{tabular}{lc}
        \begin{overpic}[width=0.98\textwidth]{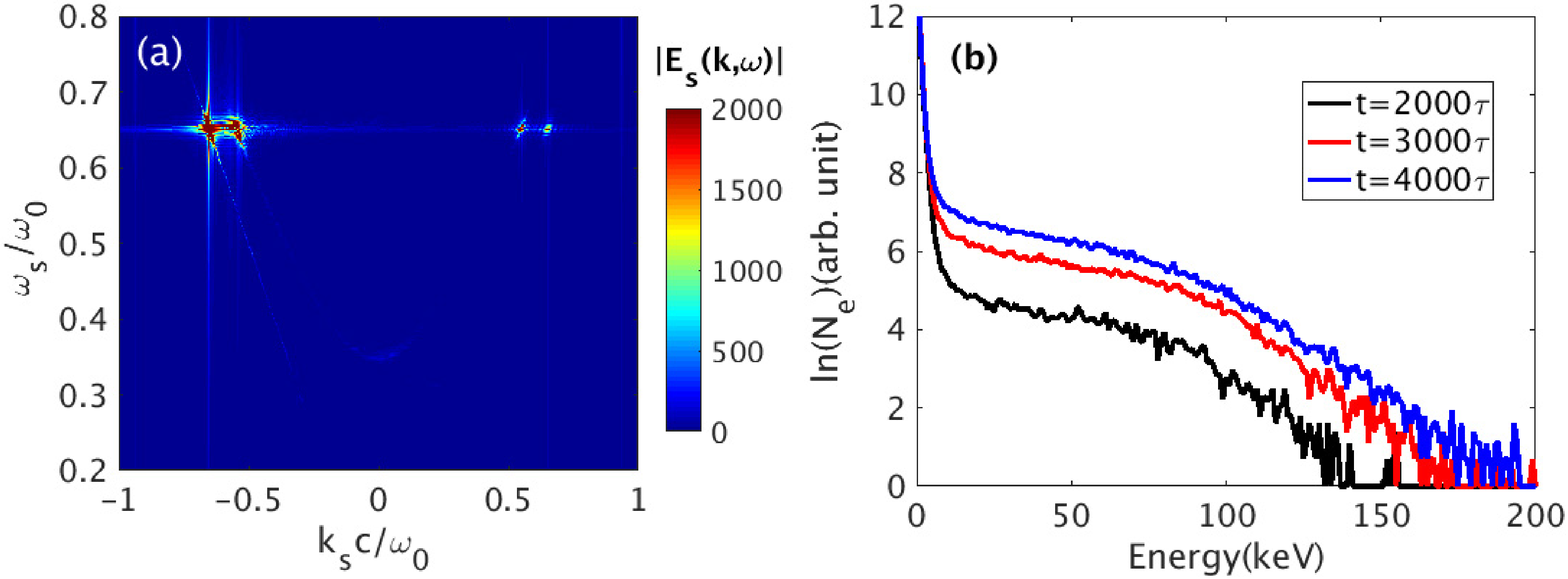}
        \end{overpic}
    \end{tabular}
\caption{ The case when the absolute SRS is absent provided $n_{min}>n_c/9$. (a) Wavenumber-frequency distributions of the scattered light in the time windows $[2001,2500]\tau$. (b) Energy distributions of electrons at different time.
    }
\end{figure}

Based on the discussion in Sec. II, a strong convective backward SRS develops in relatively high density region $n_e\ge n_c/9$ due to the large growth rate, and then the produced backscattering light induces absolute SRS mode in the region $n_e\le n_c/9$. Figure 3 illustrates the development of SRS in different stages. A strong convective backscattering SRS is developed in the time window of [1501,2000]$\tau$ as shown in Fig. 3(a), which is mainly caused by the first-order SRS scattering. The frequency of backscattering light ranges from $0.64\omega_0$ (corresponding to $n_1=0.13n_c$) to $0.72\omega_0$ (corresponding to $n_1=0.08n_c$). Note that the forward scattering mode is much weaker than the backward mode due to its relatively smaller growth rate, and therefore will not be discussed in the following. From Fig. 3(b), we find that an absolute SRS mode has been generated during [2001,2500]$\tau$. The absolute mode is identified due to $k_s\approx0$ as discussed above. The central frequency of the absolute SRS mode is around $0.32\omega_0$, which is expected according to Eq. (1). The wavenumber-frequency distribution of Langmuir wave is shown in Fig. 3(c). The absolute mode is found at $k_Lc=-0.56\omega_0$ which is consistent with the deduction from Fig. 3(b) and in agreement with Eq. (3). The backscattered light (propagates from right to left) is diagnosed at $x=10\lambda$ with its Fourier spectrum shown in Fig. 3(d). A strong spectrum around $\omega_s=0.36\omega_0$ is formed during $3001\tau-4500\tau$, which is produced via forward scattering of absolute SRS. The scattering spectrum can help us to identify the development of absolute SRS mode.

The generation mechanism for the absolute SRS mode can be well demonstrated in Fig. 3(e). One finds a strong convective mode developed at $x=580\lambda$ around $t=1900\tau$, and then it gradually spread out of the resonant region. A simple calculation indicates that the quarter-critical density for the backscattering light is around $n_e=0.105n_c$. At $t=2350\tau$, the backscattering light induces an intense absolute SRS mode near $x=320\lambda$ where the plasma density is $n_e=0.104n_c$. For the absolute SRS instability, the resonant region is stationary. One finds that the absolute instability region gradually widens to [300,345]$\lambda$, due to the enhanced bandwidth of the backscattering light. As mentioned above, a broad instability region near the quarter-critical density can greatly reduce the suppression effects of finite bandwidth of the backscattering light. After $t=2800\tau$, the absolute SRS becomes the dominant instability mode.

According to Eq. (3), the phase velocity of the Langmuir wave excited by the absolute SRS is very high. In linear stage, the temperature of most electrons are much lower than 170keV. Therefore, the absolute SRS mode can grow without Landau damping. The main damping for the growth of absolute SRS mode comes from the backscattering light developed via backward SRS. In our simulation, the phase velocity of the Langmuir wave developed by convective SRS is around $0.26c$. From Fig. 3(f), we find that the electron energy distribution is flattened around $v_x=-0.27c$ by trapped electrons at $t=2500\tau$. Note that the velocity of most electrons is less than $0.58c$. Therefore, this time period is the linear stage for the absolute SRS instability. At $t=3250\tau$, the absolute SRS mode has started evolving into the nonlinear regime, and another flatten region near $v_x=-0.6c$ is formed. Figure 3(g) presents a clear physical picture of electrons accelerated in the absolute SRS region. The maximum electron momentums are below $|p_x|\sim0.7m_ec$ at $x>350\lambda$. When the electrons propagates in to the absolute SRS region, they are accelerated up to even $|p_x|\sim1.5m_ec$.

The overall electron energy distribution has been diagnosed as shown in Fig. 3(h). Before the development of the absolute SRS, electron heating is weak at $t=2000\tau$. During 2000$\tau$ to 3000$\tau$, however, a large number of hot electrons start to be heated up to above 100keV. Afterwards when the process evolves into the nonlinear regime, lots of hot electrons can be trapped by the Langmuir wave and heated even up to 300keV, where a hot electron tail with temperature around $T_e=78$keV is generated by the absolute SRS mode at $t=4000\tau$. One finds that absolute SRS plays a leading role in the heating process after $t=2600\tau$. It has been reported that experiments in NIF demonstrate that the electrons with temperature $T_e>170$keV can cause ignition capsule preheat \cite{Dewald2010}. Therefore, the absolute SRS instability induced by convective backward SRS could be a crucial factor in the long time interactions.

To further validate the theoretical predictions on the condition of absolute SRS instability, we have performed a simulation with plasma density profile $n_e(x)=0.12[1+(x-525)/1500]n_c$ with plasma region [525,650]$\lambda$, where the minimum density is larger than $n_c/9$. The other parameters are the same to the above simulation. As discussed in Sec II, under this plasma condition, the second-order absolute SRS mode would not be triggered. As shown in Fig. 4(a), only convective SRS mode can be found in the phase-distribution of scattered light in the same time window with Fig. 3(b), i.e., no absolute SRS is developed. The absolute SRS mode is thought to be a mechanism for the production of hot electrons. As shown in Fig. 4(b), the electron energy distribution tends to be saturated after $t=3000\tau$, and the maximum electron energy is still less than 200keV at $t=4000\tau$. This simulation result further proves that hot electron production is considerably reduced if no absolute SRS mode is developed.

Based upon the above two simulations as well as the simulation with density range $[0.08,0.1]n_c$, we calculate the energy ratio of electrons heated by absolute SRS to the whole electrons with energy $>$60keV when the electron temperatures are saturated, and the result is $\sim$30\%. This calculation indicates that absolute instabilities in inhomogeneous plasma is a crucial effect on the production of hot electrons.

Now we study the absolute SRS developed in a hot plasma with movable ions. The plasma density profile is $n_e(x)=0.095[1+(x-50)/3000]n_c$, and variable $x$ ranges from 50$\lambda$ to 750$\lambda$. The initial electron and ion temperatures are respectively $T_{e0}=2$keV and $T_{i0}=0.8$keV. Even with such high temperature, the backward convective SRS has not been heavily damped. The ions are protons with a charge $Z=1$ and mass $m_i=1836m_e$.

\begin{figure}
\centering
    \begin{tabular}{lc}
        \begin{overpic}[width=0.98\textwidth]{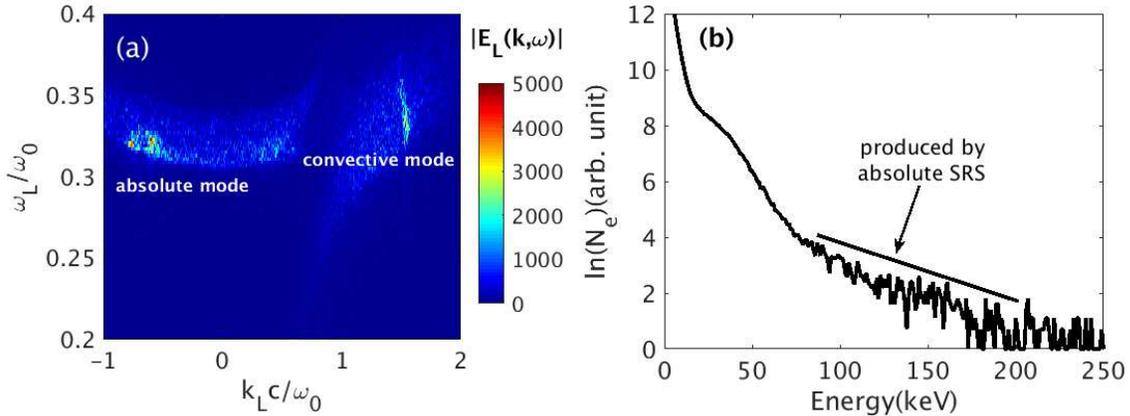}
        \end{overpic}
    \end{tabular}
\caption{ (a) 2D Fourier transform $|E_L(k,\omega)|$ of the electric field in the time windows $[1501,2000]\tau$. (b) Energy distributions of electrons at $t=4500\tau$.
}
\end{figure}

An absolute SRS mode with finite bandwidth can be found around $k_Lc=-0.58\omega_0$ in the time window [1501,2000]$\tau$ as shown in Fig. 5(a). The intensity of absolute mode is comparable to the convective backward SRS mode. One notes that absolute SRS has a relatively larger growth rate than other secondary instabilities, and it does not compete directly with the first-order convective SRS mode. A hot tail is formed in the electron energy distributions at $t=4500\tau$ as presented in Fig. 5(b). By comparing with Fig. 3(h), we find that the energy of hot electron is reduced. Besides the temperature effects, ion acoustic wave always weaken the strength of convective SRS by triggering stimulated Brillouin scattering (SBS) and other nonlinear mode \cite{Hinkel2004National}. However, the ion effects are weakened in the kinetic regime, due to the reduction of ion acoustic wave \cite{riconda,Weber2005,Zhao2017Inhibition}. In conclusion, we can find a relatively weak absolute SRS mode in the plasma with a few keV initial temperature and mobile ions.

\subsection{2D simulations for absolute instability modes}

To further validate the absolute instabilities induced via the rescatter of SRS in high dimension, we performed several 2D PIC simulations. The length of the simulation box is 550$\lambda$, where the plasma occupies a region from 25$\lambda$ to 525$\lambda$ with density profile $n_e(x)=0.0998[1+(x-25)/2000]n_c$. The width of the plasma is 20$\lambda$. The initial electron temperature is $T_{e0}=100$eV. The ions are stationary with a charge $Z=1$. Reflection boundary conditions have been set for the electrons. A p-polarized (electric field of light is parallel to the simulation plane) semi-infinite pump laser with a uniform amplitude $a_0=0.02$ is incident from the left boundary of the simulation box. We have taken 50 cells per wavelength in both transverse and longitudinal directions.

\begin{figure}
\centering
    \begin{tabular}{lc}
        \begin{overpic}[width=0.98\textwidth]{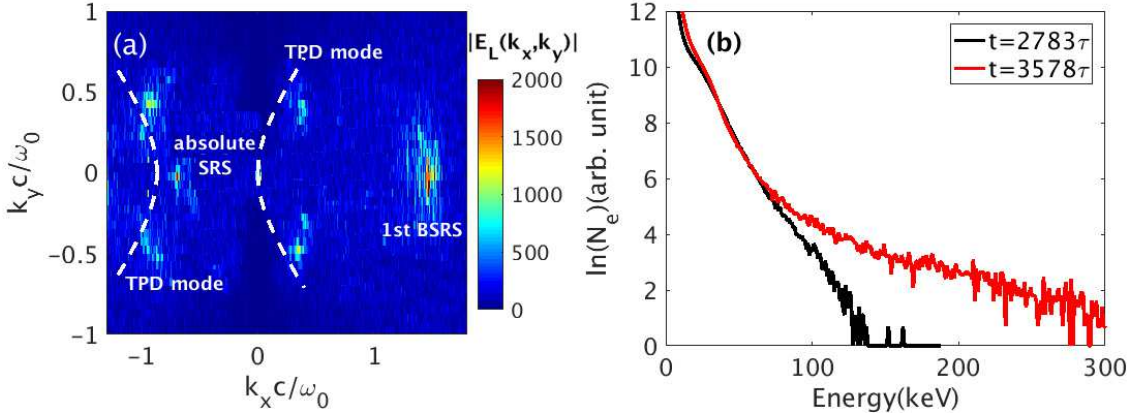}
        \end{overpic}
    \end{tabular}
\caption{ (a) Spatial Fourier transform $|E_L(k_x,k_y)|$ of the electric field at $t=3380\tau$. The white dotted line is the expected wavenumber distribution for plasmons occurring in the TPD instability. (b) Energy distributions of electrons at different time. The incident laser is p-polarization.
}
\end{figure}

As discussed above, TPD mode may also be found in the absolute instability region, therefore we have carried out simulations with p-polarized laser. The white dotted line in Fig. 6(a) is calculated from $(k_x+k_0/2)^2-k_y^2=(k_0/2)^2$ \cite{Vu2010The}, which is the expected wavenumber distribution for plasmons occurring in the induced TPD instability without considering temperature and bandwidth effects. Moreover, an absolute SRS spectrum can be found around $k_xc\sim0.6\omega_0$, which confirms that the absolute instability is not only a one-dimensional effect. A hot-electron tail is formed at $t=3578\tau$ as shown in Fig. 6(b), which further prove that the absolute instabilities induced via SRS rescattering do contribute significantly to the hot electron productions.

\section{Simulations for absolute SRS mode via the third-order rescattering of SRS}

The above simulations are calculated for absolute SRS mode via the second-order rescattering of SRS. It is possible that absolute SRS mode via the third-order scattering can be found in a large scale plasma according to our theoretical analysis. In this subsection we show simulation results to demonstrate this. An inhomogeneous plasma with density profile $n_e(x)=0.04[1+(x-25)/1000]n_c$ occupies a region from $x=25\lambda$ to $x=1275\lambda$. Note that the plasma density range covers $n_e=0.0625n_c$, and the maximum plasma density is $n_{max}=0.09n_c<n_c/9$. The initial electron temperature is $T_{e0}=100$eV. The ions are stationary with a charge $Z=1$.

\begin{figure}
\centering
    \begin{tabular}{lc}
        \begin{overpic}[width=0.98\textwidth]{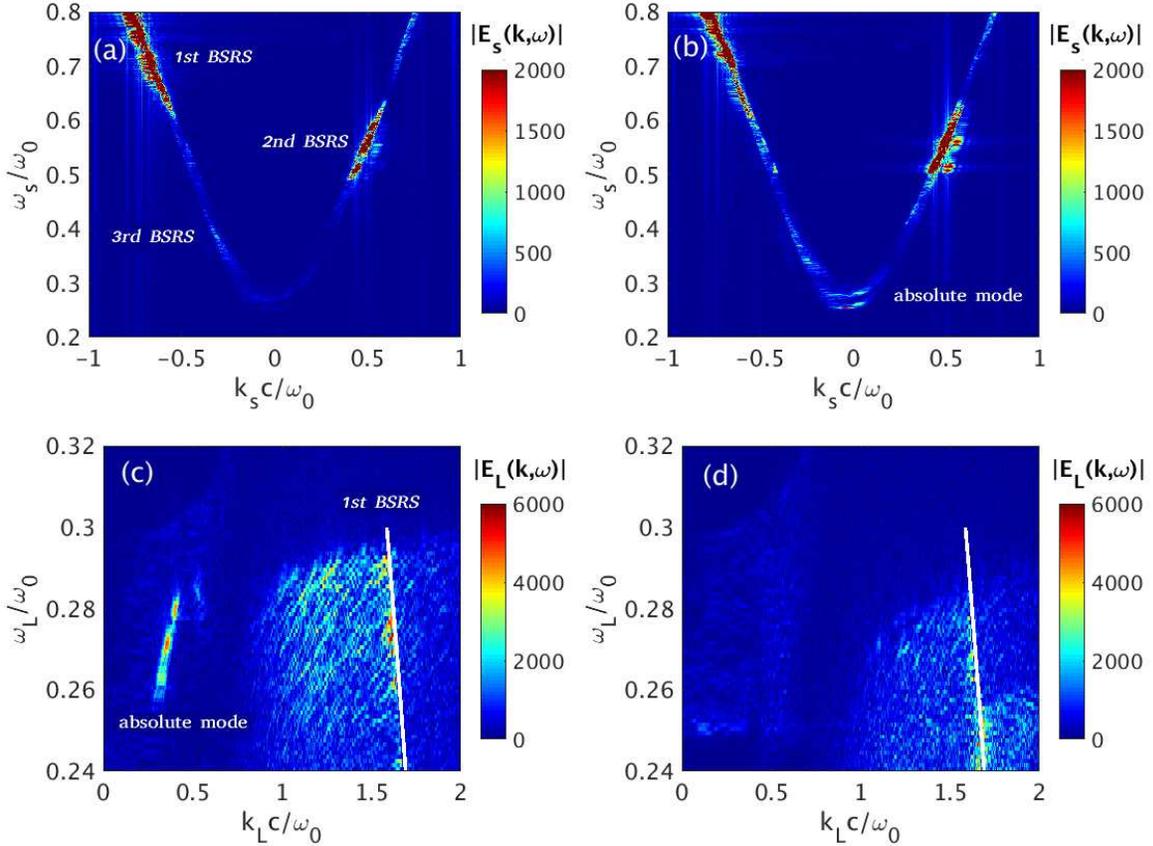}
        \end{overpic}
    \end{tabular}
\caption{ PIC simulation results for the development of the absolute SRS via the third-order scattering. (a)-(c) Simulation results for the plasma with the inhomogeneous plasma density range [0.04,0.09]$n_c$. (a) and (b) show the 2D Fourier transform $|E_s(k,\omega)|$ of the scattered light $E_s(x,t)$ in the time windows $[3001,4000]\tau$ and $[4001,5000]\tau$, respectively. (c) 2D Fourier transform $|E_L(k,\omega)|$ of the electric field in the time windows $[5001,6000]\tau$. The white line denotes the linear resonant region for convective backscattering SRS. (d) 2D Fourier transform $|E_L(k,\omega)|$ of the electric field in the time windows $[5001,6000]\tau$, when the plasma density profile is limited to the range of [0.0625,0.09]$n_c$.
}
\end{figure}

Figure 7 illustrates the evolution of the scattering waves and Langmuir waves in the wavenumber-frequency space. Figure 7(a) shows clearly the cascaded scattering has developed, including the first-order scattering, the second-order scattering, and even the third-order scattering. It also suggests that a third-order scattering is the convective backward SRS mode, which is generated around $t=4000\tau$ at a very low intensity. Some time later, a relatively strong absolute mode can be found in the time window [4001,5000]$\tau$ as shown in Fig. 7(b). The absolute instability mode can be clearly distinguished from the convective one in the spectrum distribution. One notes that the absolute mode is developed in the density region ranging from 0.066$n_c$ to 0.077$n_c$ in this period.

Figure 7(c) shows the distribution of the Langmuir wave in $(k_L,\omega_L)$ space. The white line marks the linear resonant region for the convective backward SRS obtained from $k_{L}c=\sqrt{\omega_0^2-\omega_{L}^2}+\sqrt{\omega_0^2-2\omega_0\omega_{L}}$. Broad wave number spectrum indicates that the convective modes spread in space with strong frequency shift \cite{winjum2007relative}. Moreover, the generation of the third-order backward SRS and the increasing of electron temperature can also lead to the wave number shift. The absolute mode around $k_L=0.4\omega_0/c$ with bandwidth 0.022$\omega_0$ demonstrates that the absolute SRS instability region is [0.067,0.079]$n_c$, which is the same as the range inferred from Fig. 7(b). One notes that the intensity of the absolute SRS mode becomes comparable to the first-order backward SRS at 6000$\tau$. When the absolute SRS mode induced by third-order rescattering of SRS suffers Landau damping, then the Langmuir wave will transfer energy to the hot electrons.

As comparison, we have performed another set of simulation with a plasma density profile limited to the range [0.0625,0.09]$n_c$ with other parameters unchanged. Based on the discussion in Sec. II, the development of the absolute SRS mode via the third-order scattering needs a second-order backward SRS region $n_e<0.0625n_c$ to generate backscattering lights. Without this density region, only a convective mode can be found and no third-order absolute SRS can be developed. This is verified in Fig. 7(d).

\section{Summary}

In summary, we have theoretically and numerically studied the mechanism of the absolute instabilities induced by cascading SRS processes under indirect-drive conditions, where a large scale of nonuniform plasma is expected. Theoretical analysis indicates that the backscattering light from convective SRS can generate strong absolute instabilities in a plasma within $n_e<1/9n_c$ via the second-order SRS rescattering. The absolute SRS mode develops a Langmuir wave with a high phase velocity about 0.58$c$, corresponding to hot electron energy around 170keV. In the linear stage of the absolute SRS instability, the produced hot electron energy is much lower than 100keV, and the absolute SRS mode can grow without Landau damping. When the interactions evolve into the nonlinear regime, a large number of electrons are trapped by the Langmuir wave and are heated up to 300keV. It is further demonstrated that the absolute SRS instability can be developed via the third-order backward SRS when the inhomogeneous plasma is even larger and the interaction time is long over a few picoseconds. A TPD mode can be found in the absolute instability region, which also produces hot electrons with different energies. The absolute instabilities due to rescatter of SRS may play a significant role for the capsule preheating in the experiments of inertial confined fusion, where a large inhomogeneous plasma is often involved.

\section{Acknowledgement}

The authors acknowledge useful discussions with C. Z. Xiao, R. Yan and C. Ren. This work was supported by National Science Foundation of China (Grant No. 11775144 and 1172109).


\section*{References}
\begin{thebibliography}{10}
\bibitem{kruer1988physics} Kruer W L 1988 The physics of laser plasma interactions (Addison-Wesley, New York)
\bibitem{liu1974raman} Liu C S \etal. 1974 {\it Phys. Fluids} {\bf 17}, 1211
\bibitem{Pesme2002Laser} Pesme D \etal. 2002 {\it Plasma Phys. Control. Fusion} {\bf 44}, B53
\bibitem{Montgomery2016Two} Montgomery D S \etal. 2016 {\it Phys. Plasmas} {\bf 23}, 055601
\bibitem{Betti2016Inertial} Betti R and Hurricane O A 2016 {\it Nature Phys.} {\bf 12}, 435
\bibitem{Lindl2014Review} Lindl J \etal. 2014 {\it Phys. Plasmas} {\bf 21}, 020501
\bibitem{Hurricane2014Fuel} Hurricane O A \etal. 2014 {\it Nature} {\bf 506}, 343
\bibitem{Campbell2017Laser} Campbell E M \etal. 2017 {\it Matter Radiat. Extrem.} {\bf 2}, 37
\bibitem{He2016A} He X T \etal. 2016 {\it Phys. Plasmas} {\bf 23}, 082706
\bibitem{Smalyuk2008Role} Smalyuk V A \etal. 2008 \PRL {\bf 100}, 185005
\bibitem{Sangster2008High} Sangster T C \etal. 2008 \PRL {\bf 100}, 185006
\bibitem{Batani2014} Batani D \etal. 2014 {\it Nucl. Fusion} {\bf 54}, 054009
\bibitem{Dewald2010} Dewald E L \etal. 2010 {\it Rev. Sci. Instrum.} {\bf 81}, 10D938
\bibitem{Regan2010Suprathermal} Regan S P \etal. 2010 {\it Phys. Plasmas} {\bf 17}, 055503
\bibitem{Glenzer2004Progress} Glenzer S H \etal. 2004 {\it Nucl. Fusion} {\bf 44}, S185
\bibitem{Strozzi2008} Strozzi D J \etal. 2008 {\it Phys. Plasmas} {\bf 15}, 102703
\bibitem{winjum2013anomalously} Winjum B J \etal. 2013 \PRL {\bf 110}, 165001
\bibitem{Kruer1980Raman} Kruer W L \etal. 1980 {\it Phys. Fluids} {\bf 23}, 1326
\bibitem{Hinkel2004National} Hinkel D E \etal. 2004 {\it Phys. Plasmas} {\bf 11}, 1128
\bibitem{Mima2001Stimulated} Mima K \etal. 2001 {\it Phys. Plasmas} {\bf 8}, 2349
\bibitem{Klimo2010Particle} Klimo O \etal. 2010 {\it Plasma Phys. Control. Fusion} {\bf 52}, 055013
\bibitem{White1974Absolute} White B \etal. 1974 {\it Nucl. Fusion} {\bf 14}, 45
\bibitem{myatt2014multiple} Myatt J F \etal. 2014 {\it Phys. Plasmas} {\bf 21}, 055501
\bibitem{Langdon2002Nonlinear} Langdon A B and Hinkel D E 2002 \PRL {\bf 89}, 015003
\bibitem{Fern2000Observed} Fernandez J C \etal. 2000 {\it Phys. Plasmas} {\bf 7}, 3743
\bibitem{YaoZ2017Effective} Zhao Y \etal. 2017 {\it Phys. Plasmas} {\bf 24}, 112102
\bibitem{Liberal2017} Liberal I \etal. 2017 {\it Nature Photon.} {\bf 11}, 149
\bibitem{Ahmedm2014} Ahmed M M \etal. 2014 {\it Nature Commun.} {\bf 5}, 5638
\bibitem{chen2008development} Chen M \etal. 2008 {\it Chin. J. Comput. Phys.} {\bf 25}, 43 (in Chinese)
\bibitem{riconda} Riconda C \etal. 2005 \PRL {\bf 94}, 055003
\bibitem{Weber2005} Weber S, Riconda C, and Tikhonchuk V T 2005 \PRL {\bf 94}, 055005
\bibitem{Zhao2017Inhibition} Zhao Y \etal. 2017 {\it Phys. Plasmas} {\bf 24}, 092116
\bibitem{Vu2010The} Vu H X \etal. 2010 {\it Phys. Plasmas} {\bf 17}, 072701
\bibitem{winjum2007relative} Winjum B J \etal. 2007 {\it Phys. Plasmas} {\bf 14}, 102104
\end{thebibliography}
\end{document}